\documentclass[aps,showpacs,11pt,tightenlines,preprint,nofootinbib]{revtex4-1}
\pdfoutput=1
\usepackage{amsmath}
\usepackage{amssymb}
\usepackage{graphicx}
\usepackage{hyperref}
\usepackage[utf8]{inputenc}
\usepackage{subfig}
\usepackage{url}
\usepackage[table]{xcolor}
\usepackage{xspace}

\newcommand{\HEPfit}{\texttt{HEPfit}\xspace}

\hypersetup{pdfstartview=FitV,colorlinks=true,linkcolor=blue,citecolor=red,filecolor=black,urlcolor=blue}

\begin{document}

\title{New theoretical constraints on scalar color octet models}

\author{Li Cheng}
\email{clmamuphy@sina.cn}

\affiliation{Institute of Applied Physics and Computational Mathematics, Beijing 100094, China}

\author{Otto Eberhardt}
\email{otto.eberhardt@ific.uv.es}

\affiliation{Instituto de F\'{i}sica Corpuscular, Universitat de Val\`{e}ncia -- CSIC, Apt.~Correus 22085, E-46071 Val\`{e}ncia, Spain}

\author{Christopher W. Murphy}
\email{cmurphy@bnl.gov}

\affiliation{Department of Physics, Brookhaven National Laboratory, Upton, New York, 11973, USA}

\date{\today}

\begin{abstract}
We study theoretical constraints on a model whose scalar sector contains one color octet
and one or two color singlet $SU(2)_L$ doublets.
Using the unitarity of the theory, we constrain the parameters of the scalar potential for the first time at next-to-leading order in perturbation theory.
We also derive new conditions guaranteeing the stability of the potential. We use the HEPfit package to single out the viable parameter regions at the electroweak scale and test the stability of the renormalization group evolution up to the multi-TeV region.
Additionally, we set upper limits on the scalar mass splittings. All results are given for both cases, with and without a second scalar color singlet.
\end{abstract}

\maketitle

\section{Introduction}
\label{sec:intro}

The discovery of the $125$~GeV Higgs boson at the LHC~\cite{Aad:2012tfa, Chatrchyan:2012xdj} exemplifies the success of the Standard Model (SM).
Considering the experimental precision, possibilities of physics beyond the SM (BSM) are not excluded yet, however, given the null results of the direct searches, any BSM physics that is discovered is likely to be more exotic in nature.
It is therefore important to determine to what extent different BSM scenarios are still viable.
Inspired by the supersymmetric models, the Two-Higgs-Doublet Model (2HDM), adding another Higgs doublet, is one of the simplest and most commonly studied extensions of the SM.
Manohar and Wise (MW), starting from the principal of Minimal Flavor Violation (MFV), proposed an alternative model~\cite{Manohar:2006ga}. It follows from MFV that the scalar sector can have only two representations, color-singlet and color-octet. Therefore, they construct an extension of the SM by adding a color-octet electroweak doublet scalar.
The phenomenology of the model has been studied in detail~\cite{Gresham:2007ri, Gerbush:2007fe, Burgess:2009wm, He:2011ti, Dobrescu:2011aa, Bai:2011aa, Arnold:2011ra, Kribs:2012kz, Reece:2012gi, Cao:2013wqa, He:2013tla, Cheng:2015lsa, Martinez:2016fyd, Hayreter:2017wra}, including aspects such as production of scalars, the lower limit of the scalar masses and possible constraints on the parameter space.

Combining the above motivations, Ref.~\cite{Cheng:2016tlc, Cheng:2017tbn} recently proposed a new model containing aspects of both MW model and 2HDM. In particular, the scalar sector of the model in consideration consists of two color-singlet electroweak doublets, $\Phi_{1, 2}$, and one color-octet electroweak doublet, $S$.
The MW model is the limiting case with $\Phi_2 \to 0$, whereas the 2HDM is recovered in the limit $S \to 0$.
Due to the existence of these two limiting cases we will refer to this model as the \textit{2HDMW}.
The inclusive character of the 2HDMW model is capable of explaining new physics and meanwhile compatible with the established experimental observations. For example, it is a viable model for LHC physics in terms of $h$-signal strengths since they are not necessarily affected at tree-level. It is also suggested that the 2HDMW can emerge naturally from GUT theories \cite{Georgi:1979df,Dorsner:2006dj,Perez:2016qbo}. Therefore, the 2HDMW is one of the possible physics at the low scale upon the breaking down of the more general symmetry.  In the meantime, $CP$ violating phases are introduced to the scalar sector in its most general formulation.

Ref.~\cite{Cheng:2016tlc} investigated tree-level constraints on the 2HDMW arising from symmetries and perturbative unitarity.
A study of LHC phenomenology was also performed, and found that the color-octet scalar added to the 2HDM could produce large corrections to the one-loop couplings of the Higgs boson to two gluons or photons.
Ref.~\cite{Cheng:2017tbn} derived the one-loop beta functions for the scalar couplings in the 2HDMW, and the evolution of the renormalization group equations (RGEs) was then used to place upper limits on the parameters of the model.
Similar practices were applied in studies of the the SM~\cite{Callaway:1988ya, Sher:1988mj}, the MW model~\cite{He:2013tla}, and the 2HDM~\cite{Chakrabarty:2014aya, Chowdhury:2015yja, Ferreira:2015rha}.
The parameter space was further constrained in Ref.~\cite{Cheng:2017tbn} by requiring no Landau poles (LPs) below a certain high energy scale $\Lambda$, the scalar potential being stable and perturbative unitarity satisfied at all scales below $\Lambda$.
The perturbative unitarity constraints imposed on the model in Ref.~\cite{Cheng:2016tlc, Cheng:2017tbn} are leading order (LO), and a considerable region in the parameter space survives.

Although instuctive, the preceding studies on constraints imposed on the 2HDMW are not yet comprehensive.
It is a reasonable expectation that supplementing corrections at higher orders can result in noticeable modifications to the surviving parameter space. However, the behavior of higher order corrections is usually complicated. Whether their impact is to tighten or relax the viable ranges of couplings, there is no simple answer to that. In this paper, we utilize the generic tool provided by Ref.~\cite{Grinstein:2015rtl, Cacchio:2016qyh, Murphy:2017ojk} to explore these perturbative unitarity bounds at next-to-leading order (NLO) and firstly impose them on color-octet scalar. On the other hand, the positivity conditions are only known for 2HDM. With additional color-octet taking into play, one should reconsider the scalar potential as a whole and secure the existence of the global minimum. Completely solving this problem is extremely challenging. This paper is also the first work to expand the set of positivity conditions to both MW and 2HDMW models.

More generally this work focuses on theoretical constraints on the 2HDMW.
An investigation of experimental bounds on the model is saved for future work.
The rest of the paper is organized as follows:
The 2HDMW model is defined in Section~\ref{sec:model}.
The theoretical constraints are explained in Sec.~\ref{sec:constraints}.
Following that our results for the surviving parameter space are presented in Sec.~\ref{sec:results}. 
Concluding remarks are given in Sec.~\ref{sec:conclusions}.

\section{The model}
\label{sec:model}

As stated above, the scalar sector of the model consists of two color-singlet electroweak doublets $\Phi_{1, 2}$, and one color-octet electroweak doublet $S$.
The most general renormalizable potential of the scalar sector is \cite{Cheng:2016tlc, Branco:2011iw}:

\begin{align}
 V_{\text{\tiny{gen}}}
 & = m_{11}^2\Phi_1^\dagger\Phi_1^{\phantom{\dagger}}
   + m_{22}^2\Phi_2^\dagger\Phi_2^{\phantom{\dagger}}
   - m_{12}^2 \left( \Phi_1^\dagger\Phi_2^{\phantom{\dagger}}
              +\Phi_2^\dagger\Phi_1^{\phantom{\dagger}}\right)
   + \tfrac12 \lambda_1\left(\Phi_1^\dagger\Phi_1^{\phantom{\dagger}}\right)^2
   + \tfrac12 \lambda_2\left(\Phi_2^\dagger\Phi_2^{\phantom{\dagger}}\right)^2 \nonumber \\
 &\phantom{{}={}}
  + \lambda_3 \left(\Phi_1^\dagger\Phi_1^{\phantom{\dagger}}\right)
            \left(\Phi_2^\dagger\Phi_2^{\phantom{\dagger}}\right)
  + \lambda_4 \left(\Phi_1^\dagger\Phi_2^{\phantom{\dagger}}\right)
            \left(\Phi_2^\dagger\Phi_1^{\phantom{\dagger}}\right)
  + \tfrac12 \left[ \lambda_5 \left(\Phi_1^\dagger\Phi_2^{\phantom{\dagger}}\right)^2
                      + {\rm h.c.} \right] \nonumber \\
 &\phantom{{}={}}
  + \left[ \lambda_6 \left( \Phi_1^\dagger\Phi_1^{\phantom{\dagger}} \right) \left( \Phi_1^\dagger\Phi_2^{\phantom{\dagger}} \right) 
   +\lambda_7 \left( \Phi_2^\dagger\Phi_2^{\phantom{\dagger}} \right) \left( \Phi_1^\dagger\Phi_2^{\phantom{\dagger}} \right) 
                      + {\rm h.c.} \right] \nonumber \\
 &\phantom{{}={}}
  + 2 m_S^2 {\rm Tr}\left(S^{\dagger i} S^{\phantom{\dagger}}_i\right)
  + \mu_1 {\rm Tr}\left(S^{\dagger i} S^{\phantom{\dagger}}_i S^{\dagger j} S^{\phantom{\dagger}}_j\right) 
  + \mu_2 {\rm Tr}\left(S^{\dagger i} S^{\phantom{\dagger}}_j S^{\dagger j} S^{\phantom{\dagger}}_i\right) 
  + \mu_3 {\rm Tr}\left(S^{\dagger i} S^{\phantom{\dagger}}_i\right) \left(S^{\dagger j} S^{\phantom{\dagger}}_j\right) \nonumber \\
 &\phantom{{}={}}
  + \mu_4 {\rm Tr}\left(S^{\dagger i} S^{\phantom{\dagger}}_j\right) \left(S^{\dagger j} S^{\phantom{\dagger}}_i\right) 
  + \mu_5 {\rm Tr}\left(S^{\phantom{\dagger}}_i S^{\phantom{\dagger}}_j\right) \left(S^{\dagger i} S^{\dagger j}\right) 
  + \mu_6 {\rm Tr}\left(S^{\phantom{\dagger}}_i S^{\phantom{\dagger}}_j S^{\dagger j} S^{\dagger i}\right) \nonumber \\
 &\phantom{{}={}}
  + \nu_1 \Phi_1^{\dagger i}\Phi_{1i}^{\phantom{\dagger}} {\rm Tr}\left(S^{\dagger j} S^{\phantom{\dagger}}_j\right) 
  + \nu_2 \Phi_1^{\dagger i}\Phi_{1j}^{\phantom{\dagger}} {\rm Tr}\left(S^{\dagger j} S^{\phantom{\dagger}}_i\right) \nonumber \\
 &\phantom{{}={}}
 +\left[ \nu_3 \Phi_1^{\dagger i}\Phi_1^{\dagger j} {\rm Tr}\left(S^{\phantom{\dagger}}_i S^{\phantom{\dagger}}_j\right) 
           +\nu_4 \Phi_1^{\dagger i} {\rm Tr}\left(S^{\dagger j} S^{\phantom{\dagger}}_j S^{\phantom{\dagger}}_i\right) 
           +\nu_5 \Phi_1^{\dagger i} {\rm Tr}\left(S^{\dagger j} S^{\phantom{\dagger}}_i S^{\phantom{\dagger}}_j\right) 
           + {\rm h.c.}\right] \nonumber \\
 &\phantom{{}={}}
  + \omega_1 \Phi_2^{\dagger i}\Phi_{2i}^{\phantom{\dagger}} {\rm Tr}\left(S^{\dagger j} S^{\phantom{\dagger}}_j\right) 
  + \omega_2 \Phi_2^{\dagger i}\Phi_{2j}^{\phantom{\dagger}} {\rm Tr}\left(S^{\dagger j} S^{\phantom{\dagger}}_i\right) \nonumber \\
 &\phantom{{}={}}
 +\left[ \omega_3 \Phi_2^{\dagger i}\Phi_2^{\dagger j} {\rm Tr}\left(S^{\phantom{\dagger}}_i S^{\phantom{\dagger}}_j\right) 
           +\omega_4 \Phi_2^{\dagger i} {\rm Tr}\left(S^{\dagger j} S^{\phantom{\dagger}}_j S^{\phantom{\dagger}}_i\right) 
           +\omega_5 \Phi_2^{\dagger i} {\rm Tr}\left(S^{\dagger j} S^{\phantom{\dagger}}_i S^{\phantom{\dagger}}_j\right) 
           +{\rm h.c.}\right] \nonumber \\
 &\phantom{{}={}}
 +\left[ \kappa_1 \Phi_1^{\dagger i} \Phi_{2i}^{\phantom{\dagger}} {\rm Tr}\left(S^{\dagger j} S^{\phantom{\dagger}}_j\right) 
           +\kappa_2 \Phi_1^{\dagger i} \Phi_{2j}^{\phantom{\dagger}} {\rm Tr}\left(S^{\dagger j} S^{\phantom{\dagger}}_i\right) 
           +\kappa_3 \Phi_1^{\dagger i} \Phi_2^{\dagger j} {\rm Tr}\left(S^{\phantom{\dagger}}_j S^{\phantom{\dagger}}_i\right)
           +{\rm h.c.}\right]
           \label{eq:genpot}
\end{align}

All interactions between $S$, $\Phi_1$, and $\Phi_2$ and the self-interactions are included.
In Eq.~\eqref{eq:genpot}, we use $i,j$ as $SU(2)$ indices; the notation $S_i = S_i^A T^A$, where $A$ is color index. The trace is taken over the color indices.

The physical parameters of this model are the masses of the $\Phi_1$ and $\Phi_2$ fields, which we denote like in the 2HDM as $m_h$, $m_H$ and $m_A$ for the neutral bosons and as $m_{H^\pm}$ for the charged Higgs particles, as well as the octet masses $m_R$, $m_I$ and $m_{S^\pm}$ for the neutral scalar, the neutral pseudoscalar and the charged octet scalar of the MW model. Moreover, we will call the two angles of the diagonalization of the mass matrix in the 2HDM sector $\alpha$ and $\beta$, according to the convention in the literature.

We apply the following conditions to reduce the number of the parameters in the scalar potential and the Yukawa potential, defined below in Eq.~\eqref{eq:pot} and Eq.~\eqref{eq:yukawa}, respectively.

\begin{itemize}
\item We restrict the 2HDM sector to be $CP$-conserving.
\item Custodial symmetry~\cite{Sikivie:1980hm, Pomarol:1993mu, Grzadkowski:2010dj}: We adopt the less restrictive method discussed in~\cite{Cheng:2016tlc}.
The mass degeneracies $m_{H^\pm} = m_A$ and $m_{S^\pm} = m_I$ result from custodial symmetry.
\item We impose a $\mathbb{Z}_2$ symmetry, which is only softly broken by quadratic terms.
This prevents tree level flavor changing neutral currents (FCNCs), and further reduces the number of free parameters.
The charge assignments we consider are given in Table~\ref{tab:types}.
Note that the original MW paper was motivated by the principle of minimal flavor violation~\cite{Chivukula:1987py, DAmbrosio:2002vsn}.
This is in contrast with our approach of imposing $\mathbb{Z}_2$ symmetry, which is motivated by the practicality of reducing the number of parameters in the scalar potential while still maintaining some ability to generate flavor effects.
\end{itemize}

\begin{table}
\centering
 \begin{tabular}{| c || c | c | c | c | c | c |}
 \hline 
 & $\Phi_1$ & $\Phi_2$ & $S$ & $U_R$ & $D_R$ & $Q_L$  \\ \hline 
Type I & $-$ & $+$ & $-$ & $-$ & $-$ & $+$  \\ 
Type IIu & $-$ & $+$ & $-$ & $-$ & $+$ & $+$  \\
Type IId & $-$ & $+$ & $+$ & $-$ & $+$ & $+$  \\ \hline 
 \end{tabular}
  \caption{$\mathbb{Z}_2$ charge assignments in the 2HDMW that forbid tree level FCNCs. In type IIu (IId) the color-octet scalar $S$ only interacts with up-type (down-type) quarks.}
  \label{tab:types}
\end{table}

The scalar potential of the model with the aforementioned constraints imposed reads 
\begin{align}
 V_{\text{\tiny{fit}}}
 & = m_{11}^2\Phi_1^\dagger\Phi_1^{\phantom{\dagger}}
   + m_{22}^2\Phi_2^\dagger\Phi_2^{\phantom{\dagger}}
   - m_{12}^2 \left( \Phi_1^\dagger\Phi_2^{\phantom{\dagger}}
              +\Phi_2^\dagger\Phi_1^{\phantom{\dagger}}\right)
   + \tfrac12 \lambda_1\left(\Phi_1^\dagger\Phi_1^{\phantom{\dagger}}\right)^2
   + \tfrac12 \lambda_2\left(\Phi_2^\dagger\Phi_2^{\phantom{\dagger}}\right)^2 \nonumber \\
 &\phantom{{}={}}
  + \lambda_3 \left(\Phi_1^\dagger\Phi_1^{\phantom{\dagger}}\right)
            \left(\Phi_2^\dagger\Phi_2^{\phantom{\dagger}}\right)
  + \tfrac12 \lambda_4 \left[ \left(\Phi_1^\dagger\Phi_2^{\phantom{\dagger}}\right)
                      + \left(\Phi_2^\dagger\Phi_1^{\phantom{\dagger}}\right) \right]^2 \nonumber \\
 &\phantom{{}={}}
  + 2 m_S^2 {\rm Tr}\left(S^{\dagger i} S^{\phantom{\dagger}}_i\right)
  + \mu_1 \left[ {\rm Tr}\left(S^{\dagger i} S^{\phantom{\dagger}}_i S^{\dagger j} S^{\phantom{\dagger}}_j\right) 
                        +{\rm Tr}\left(S^{\dagger i} S^{\phantom{\dagger}}_j S^{\dagger j} S^{\phantom{\dagger}}_i\right)  
                        +2 {\rm Tr}\left(S^{\phantom{\dagger}}_i S^{\phantom{\dagger}}_j S^{\dagger j} S^{\dagger i}\right) \right] \nonumber \\
 &\phantom{{}={}}
  + \mu_3 {\rm Tr}\left(S^{\dagger i} S^{\phantom{\dagger}}_i\right) \left(S^{\dagger j} S^{\phantom{\dagger}}_j\right) 
  + \mu_4 \left[ {\rm Tr}\left(S^{\dagger i} S^{\phantom{\dagger}}_j\right) \left(S^{\dagger j} S^{\phantom{\dagger}}_i\right) 
                        + {\rm Tr}\left(S^{\phantom{\dagger}}_i S^{\phantom{\dagger}}_j\right) \left(S^{\dagger i} S^{\dagger j}\right) \right] \nonumber \\
 &\phantom{{}={}}
  + \nu_1 \Phi_1^{\dagger i}\Phi_{1i}^{\phantom{\dagger}} {\rm Tr}\left(S^{\dagger j} S^{\phantom{\dagger}}_j\right) 
  +\tfrac12 \nu_2 \left[ \Phi_1^{\dagger i}\Phi_{1j}^{\phantom{\dagger}} {\rm Tr}\left(S^{\dagger j} S^{\phantom{\dagger}}_i\right)
			         +\Phi_1^{\dagger i}\Phi_1^{\dagger j} {\rm Tr}\left(S^{\phantom{\dagger}}_i S^{\phantom{\dagger}}_j\right) 
                                  + {\rm h.c.}\right] \nonumber \\
 &\phantom{{}={}}
 +\nu_4 \left[ \Phi_1^{\dagger i} {\rm Tr}\left(S^{\dagger j} S^{\phantom{\dagger}}_j S^{\phantom{\dagger}}_i\right) 
           +\Phi_1^{\dagger i} {\rm Tr}\left(S^{\dagger j} S^{\phantom{\dagger}}_i S^{\phantom{\dagger}}_j\right) 
           + {\rm h.c.}\right] \nonumber \\
 &\phantom{{}={}}
  + \omega_1 \Phi_2^{\dagger i}\Phi_{2i}^{\phantom{\dagger}} {\rm Tr}\left(S^{\dagger j} S^{\phantom{\dagger}}_j\right) 
  +\tfrac12 \omega_2 \left[ \Phi_2^{\dagger i}\Phi_{2j}^{\phantom{\dagger}} {\rm Tr}\left(S^{\dagger j} S^{\phantom{\dagger}}_i\right)
			         +\Phi_2^{\dagger i}\Phi_2^{\dagger j} {\rm Tr}\left(S^{\phantom{\dagger}}_i S^{\phantom{\dagger}}_j\right) 
                                  + {\rm h.c.}\right] \nonumber \\
 &\phantom{{}={}}
 +\omega_4 \left[ \Phi_2^{\dagger i} {\rm Tr}\left(S^{\dagger j} S^{\phantom{\dagger}}_j S^{\phantom{\dagger}}_i\right) 
           +\Phi_2^{\dagger i} {\rm Tr}\left(S^{\dagger j} S^{\phantom{\dagger}}_i S^{\phantom{\dagger}}_j\right) 
           + {\rm h.c.}\right] ,
           \label{eq:pot}
\end{align}
where $\omega_4 = 0$ in the Type I and the Type IIu 2HDMW and $\nu_4 = 0$ in the Type IId 2HDMW, leaving us with four massive and twelve massless parameters.
The masses of scalars and their mixing angles are obtained by diagonalizing the mass matrices of this model; 
the expressions of those physical parameters were presented in Eq.~(6) and (7) in Ref.~\cite{Cheng:2016tlc}.
For an overview over all assumptions, a comparison with the limiting cases of the 2HDM and the MW model and an account of the free parameters, we refer to table \ref{tab:modeloverview}.

The general Yukawa potential of the 2HDMW in the flavor eigenstate basis is given by
\begin{eqnarray}
L_{Y} &=  \left(- \eta_1^D {\left( Y_D \right)^a}_b {\bar D}_{R, a }\Phi_1^\dag Q_L^b 
- \eta_2^D {\left( Y_D \right)^a}_b {\bar D}_{R, a } \Phi_2^\dag Q_L^b  - \eta_1^U {\left( Y_U \right)^a}_b {\bar U}_{R, a}{\tilde \Phi}_1^\dag Q_L^b  \right. \nonumber\\
&-\left. \eta_S^D {\left( Y_D \right)^a}_b {\bar D}_{R, a}S^\dag Q_L^b  -
\eta_S^U {\left( Y_U \right)^a}_b {\bar U}_{R, a }{\tilde S}^\dag Q_L^b  \right) + {\rm h.c.} ,
\label{eq:yukawa}
\end{eqnarray}
where the $\eta_i$ are complex constants.\footnote{Note that contrary to the 2HDM convention, up-type quarks do not couple to $\Phi_2$ in our notation.}
In the Type I 2HDMW we have $\eta_2^D \equiv 0$ and in Type IIu (IId) $\eta_1^D\equiv 0$ and $\eta_S^D\equiv 0$ ($\eta_S^U\equiv 0$).
We use the convention ${\tilde H}_i = \varepsilon_{ij} H_j^*$, where $H=\Phi_{1,2},S$, and $a,b$ are flavor indices.

\begin{table}
{\centering
\begin{tabular}{l|cc|cc|cc}
& MW & dof. & 2HDM & dof. & 2HDMW & dof.\\
\hline
General & -- & (16) & -- & (13) & -- & (42)\\
\hline
$CP$ conservation & ${\rm Im}[\nu_{i+2}]=0$ & (14) & ${\rm Im}[m_{12}^2]=0$, & (10) & ${\rm Im}[m_{12}^2]={\rm Im}[\lambda_{i+4}]=0$, & (30)\\
 & & & ${\rm Im}[\lambda_{i+4}]=0$ & & ${\rm Im}[\nu_{i+2}]={\rm Im}[\omega_{i+2}]={\rm Im}[\kappa_i]=0$ &\\
\hline
Custodial & $\mu_1=\mu_2=\frac12 \mu_6$, & (9) & ${\rm Im}[m_{12}^2]=0$, & (9) & ${\rm Im}[m_{12}^2]={\rm Im}[\lambda_{i+4}]=0$, & (24)\\
symmetry & $\mu_4=\mu_5$, & & ${\rm Im}[\lambda_{i+4}]=0$, & & $\lambda_4=\lambda_5$, &\\
case 1 & $\nu_2=2\nu_3$, & & $\lambda_4=\lambda_5$ & & $\mu_1=\mu_2=\frac12 \mu_6$, $\mu_4=\mu_5$, &\\
of Ref.\cite{Cheng:2016tlc} & $\nu_4=\nu_5^*$ & & & & $\nu_2=2\nu_3$, $\nu_4=\nu_5^*$, &\\
 & & & & & $\omega_2=2\omega_3$, $\omega_4=\omega_5^*$, &\\
 & & & & & $\kappa_2=\kappa_3$ &\\
\hline
$\mathbb{Z}_2$ symmetry & -- & (16) & $\lambda_6=\lambda_7=0$ & (9) & $\omega_4=\omega_5=\lambda_6=\lambda_7=0$, & (28)\\
I/IIu &&&&& $\kappa_i=0$ &\\
\hline
$\mathbb{Z}_2$ symmetry & $\nu_4=\nu_5=0$ & (12) & $\lambda_6=\lambda_7=0$ & (9) & $\nu_4=\nu_5=\lambda_6=\lambda_7=0$, & (28)\\
IId &&&&& $\kappa_i=0$ &\\
\hline
Everything I/IIu & & (9) & & (7) & & (16)\\
\hline
Everything IId & & (8) & & (7) & & (16)\\
\hline
\end{tabular}
}
\caption{Overview over different model assumptions and their implementation and the number of free parameters (``dof.'') in the corresponding scalar potentials. The index $i$ is running from $1$ to $3$. The last two lines are combinations of all assumptions and thus represent the $CP$ conserving custodial $\mathbb{Z}_2$ symmetric models used for our fits.}
\label{tab:modeloverview}
\end{table}

\section{Theory constraints}
\label{sec:constraints}

\subsection{Priors}

For our analysis we make use of the open source package \HEPfit \cite{hepfit}, which is linked to the Bayesian Analysis Toolkit \cite{Caldwell:2008fw}. Even if we will not apply experimental constraints and thus not necessarily rely on a fitting tool, we chose this set-up for the following reasons: BAT can also deal with flat likelihood distributions and \HEPfit is optimized for a fast evaluation of the constraints. The sampling covers the whole parameter space, so we cannot miss relevant regions. This is not guaranteed if we use a random scattering approach. Furthermore, the presented \HEPfit implementation of the 2HDMW as well as the MW and 2HDM limiting cases are available for everyone \cite{hepfit} and can be used in future \HEPfit studies on these models including experimental data.
For more information about \HEPfit see Refs.~\cite{deBlas:2016ojx, Cacchio:2016qyh}.\\
In our Bayesian fits we use flat priors for the 2HDMW parameters with the following ranges:
\begin{align*}
&-50<\lambda_i,\mu_i,\nu_i,\omega_i<50\\
&-2<\log_{10}(\tan\beta)<2\\
&0 \;\text{GeV}^2<m_S^2<(1000 \;\text{GeV})^2
\end{align*}
We fix $\beta-\alpha$ to $\pi/2$ in order to align the light Higgs $h$ with the SM Higgs and reproduce its signal strength values at tree-level; and we set $m_{12}^2=0$ because its value is not relevant here. Note that we do not require $m_h=125$ GeV in the 2HDMW. This constraint can always be accomplished by adjusting $m_{12}^2$. Only in the MW limiting case with $\Phi_2=0$, we impose that the SM-like Higgs has a mass of $125.18\pm0.16$ GeV \cite{Aad:2015zhl,Sirunyan:2017exp}, which results in an almost fixed $\lambda_1$, like in the SM.

\subsection{Unitarity}
\label{sec:unitarity}
The unitarity of the $S$-matrix can be used to place constraints on the parameters of a theory~\cite{Lee:1977eg} (see also~\cite{He:2013tla, Kanemura:1993hm, Horejsi:2005da, Ginzburg:2005dt, Grinstein:2015rtl, Goodsell:2018tti, Goodsell:2018fex}).
If a certain combination of parameters becomes too large, an amplitude will appear to be non-unitary at a given order in perturbation theory.
We will refer to these constraints as perturbative unitarity bounds, or just unitarity bounds for short, even though the more accurate statement is that perturbation theory is breaking down.

Considering only two-to-two scattering these constraints take the following forms at various orders in perturbation theory
\begin{align}
\label{eq:Ubounds}
&\text{LO:} \quad \left(a_j^{(0)}\right)^2 \leq \frac{1}{4}, \\
&\text{NLO:} \quad 0 \leq \left(a_j^{(0)}\right)^2  + 2 \left(a_j^{(0)}\right) \text{Re}\left(a_0^{(1)}\right) \leq \frac{1}{4} , \nonumber \\
&\text{NLO+:} \quad \left[\left(a_j^{(0)}\right) + \text{Re}\left(a_j^{(1)}\right)\right]^2 \leq \frac{1}{4},  \nonumber 
\end{align}
where $a_j^{(\ell)}$ is the contribution at the $\ell$th order in perturbation theory to the $j$th partial wave amplitude.
The NLO+ inequality includes the square of NLO correction, and thus contains some, but not all of the NNLO contributions to the partial-wave amplitude.
When considering the scattering of scalars at high energy only the $j = 0$ partial wave amplitude is important.
The matrix of partial wave amplitudes is given by
\begin{equation}
\left(\mathbf{a_0}\right)_{i,f} = \frac{1}{16 \pi s} \int_{-s}^0 \! dt \, \mathcal{M}_{i \to f}(s, t) ,
\end{equation}
and we use $a_0$ to indicate the eigenvalues of $\mathbf{a_0}$.

The two-to-two scattering matrix at tree level in the neutral, color singlet channel of the 2HDMW model was recently derived in Refs.~\cite{Cheng:2016tlc, Cheng:2017tbn}.
As the scalar potential, Eq.~\eqref{eq:genpot}, contains only quartic interaction terms, the NLO unitarity bounds can be computed approximately using the algorithm of Ref.~\cite{Murphy:2017ojk}. 
A virtue of this approach is its simplicity as it only relies on knowledge of the LO partial wave matrix and the one-loop scalar contributions to the beta functions of the theory.
This algorithm is built on previous work in Ref.~\cite{Grinstein:2015rtl, Cacchio:2016qyh}, and results for the special case of the 2HDM can be found in those references.
The NLO contribution to the eigenvalue is given by a sum of two terms
\begin{equation}
a_0^{(1)} = a_{0, \sigma}^{(1)} + a_{0, \beta}^{(1)} .
\end{equation}
The first term follows from the unitarity of the theory, and is proportional to the square of the LO eigenvalue
\begin{equation}
a_0^{(1)} = \left(i - \frac{1}{\pi}\right) \left(a_0^{(0)}\right)^2 .
\end{equation}
The second term depends on the one-loop beta functions of the theory, and can be written in terms of the well known formula for the perturbations of the eigenvalues of an eigensystem for the which exact LO solution is known
\begin{equation}
a_{0, \beta}^{(1)} = \vec{x}_{(0)}^{\top} \cdot \mathbf{a}_{0, \beta}^{(1)} \cdot \vec{x}_{(0)} ,
\end{equation}
where $\vec{x}_{(0)}$ are the LO eigenvectors and
\begin{equation}
\mathbf{a}_{0, \beta}^{(1)} = - \frac{3}{2} \left.\mathbf{a}_{0}^{(0)}\right|_{\lambda_m \to \beta_{\lambda_m}} .
\end{equation}
with $\beta_{\lambda_m}$ being the beta function associated with the coupling $\lambda_m$.
The approximation is computed at a scale where the center-of-mass energy is much greater than the other scales in the problem.\footnote{Recently, finite $m^2/s$ corrections have been studied for colorless scalar SM extensions \cite{Goodsell:2018tti,Goodsell:2018fex}.}
As such we only start enforcing the unitarity bounds for RGE scales above $ 750$~GeV$\:\approx\!\sqrt{10}\:v$, and do not impose unitarity bounds when running from the EW scale to 750~GeV.

We also enforce the smallness of higher order corrections to the partial wave amplitudes with the following constraint~\cite{Grinstein:2015rtl, Cacchio:2016qyh, Durand:1992wb, Chowdhury:2017aav}
\begin{equation}
R^{\prime} \equiv \frac{\left|a_0^{(1)}\right|}{\left|a_0^{(0)}\right|} < 1 
\end{equation}
for each eigenvalue of the partial wave matrix as long as $a_0^{(0)}>0.01$.

\subsection{Boundedness from below}
\label{sec:boundedness}

In order to have a potential which is bounded from below, we extract the positivity conditions from the generic potential \eqref{eq:genpot}, assuming only that all couplings are real. Setting all but one or two of the real scalar fields to zero we require the resulting coefficient matrix to be copositive \cite{Kannike:2016fmd}.

\begin{align}
&\mu=\mu_1 + \mu_2 + \mu_6 + 2 (\mu_3 + \mu_4 + \mu_5) > 0 \label{eq:MWfirst} \\
&\mu_1 + \mu_2 + \mu_3 + \mu_4 > 0\\
&14 (\mu_1 + \mu_2) + 5 \mu_6 + 24 (\mu_3 + \mu_4) - 3 \left| 2 (\mu_1 + \mu_2) - \mu_6 \right| > 0\\
&5 (\mu_1 + \mu_2 + \mu_6) + 6 (2\mu_3 + \mu_4 + \mu_5) - \left| \mu_1 + \mu_2 + \mu_6 \right| > 0\\
&\nu_1 + \sqrt{\lambda_1 \mu }> 0\\
&\nu_1 + \nu_2 - 2|\nu_3| + \sqrt{\lambda_1 \mu}> 0\\
&\lambda_1 + \frac14 \mu + \nu_1 + \nu_2 + 2 \nu_3 - \frac{1}{\sqrt{3}}|\nu_4+\nu_5| >0\\
&\lambda_1> 0\label{eq:MWlast}\\
&\lambda_2> 0\\
&\lambda_3+\sqrt{\lambda_1 \lambda_2}> 0\\
&\lambda_3+\lambda_4-|\lambda_5|+\sqrt{\lambda_1 \lambda_2}> 0\\
&\frac12(\lambda_1+\lambda_2)+\lambda_3+\lambda_4+\lambda_5-2|\lambda_6+\lambda_7|> 0 \label{eq:2HDMlast}\\
&\omega_1 + \sqrt{\lambda_2 \mu}> 0 \label{eq:mixfirst}\\
&\omega_1 + \omega_2 - 2|\omega_3| + \sqrt{\lambda_2 \mu}> 0\\
&\lambda_2 + \frac14 \mu + \omega_1 + \omega_2 + 2 \omega_3 - \frac{1}{\sqrt{3}}|\omega_4+\omega_5| >0\label{eq:mixlast}
\end{align}

We want to stress that these conditions are necessary but not sufficient, since we did not analyze the cases with three or more non-zero fields, leaving the $\kappa_i$ unconstrained. While the pure 2HDM inequalities \eqref{eq:MWlast} to \eqref{eq:2HDMlast} have been known before \cite{Deshpande:1977rw,Branco:2011iw}, we are not aware of such conditions in the Manohar-Wise model; that is why we derive \eqref{eq:MWfirst} to \eqref{eq:MWlast} in the most general way. Finally, \eqref{eq:mixfirst} to \eqref{eq:mixlast} only appear in the 2HDMW.

In our simplified potential $V_{\text{\tiny{fit}}}$, the positivity conditions reduce to

\begin{align}
&\mu'=4\mu_1 + 2\mu_3 + 4\mu_4 > 0, \qquad 5 \mu_1 + 3 \mu_3 + 3 \mu_4 - \left| \mu_1 \right| > 0, \\
&\nu_1 + \sqrt{\lambda_1 \mu'}> 0, \qquad \nu_1 + 2\nu_2 + \sqrt{\lambda_1 \mu'} > 0, \qquad \lambda_1 + \frac14 \mu' + \nu_1 + 2\nu_2 - \frac{2}{\sqrt{3}}|\nu_4| > 0, \nonumber \\
&\lambda_1> 0, \qquad \lambda_2> 0, \qquad \lambda_3+\sqrt{\lambda_1 \lambda_2} > 0, \qquad \lambda_3+2\lambda_4+\sqrt{\lambda_1 \lambda_2} > 0, \nonumber \\
&\omega_1 + \sqrt{\lambda_2 \mu'}> 0, \qquad \omega_1 + 2\omega_2 + \sqrt{\lambda_2 \mu'}> 0, \qquad \lambda_2 + \frac14 \mu' + \omega_1 + 2\omega_2 - \frac{2}{\sqrt{3}}|\omega_4| > 0. \nonumber
\end{align}

\subsection{Positivity of the mass squares}

Additional bounds are derived from requiring the masses of the colored scalars to be real:
 
\begin{equation}
\nu_1 c_{\beta}^2 + \omega_1 s_{\beta}^2 > - \frac{4 m_S^2}{v^2} , \quad
(\nu_1 + 2 \nu_2) c_{\beta}^2 + (\omega_1 + 2 \omega_2) s_{\beta}^2 > - \frac{4 m_S^2}{v^2} . 
\end{equation} 
with $v = \sqrt{v_1^2 + v_2^2} \approx 246$~GeV, and where $s_{\beta}$ and $c_{\beta}$ are sine and cosine of $\beta$, respectively with $\tan\beta = v_1 / v_2$.
We must have $m_S^2 > 0$ so that the vacuum preserves $SU(3)_C$.
Note that the mass splitting between the colored states is
\begin{equation}
\frac{2}{v^2} \left(m_R^2 - m_{S^{\pm}}^2\right) = \nu_2 c_{\beta}^2 + \omega_2 s_{\beta}^2 \label{eq:mRsqminusmIsq},
\end{equation} 
and $m_{S^{\pm}}^2 = m_I^2$ due to custodial symmetry.

\subsection{Renormalization group stability}
So far we only discussed theory constraints at the electroweak scale.
Assuming the validity of the model up to some higher scale imposes bounds on the parameters:
Scenarios that define a viable model at $m_Z$ could feature one (or more) quartic couplings with an unstable
behavior under the renormalization group evolution to a higher scale.
This could be due to a Landau pole, but also the boundedness-from-below criteria described in Section \ref{sec:boundedness} should be fulfilled at any scale.
Furthermore, the unitarity conditions should be applied at least above some scale, $\mu_u$, as they are computed in the limit $\mu_u\gg \sqrt{\lambda_i} v$ with $\lambda_i$ being a quartic coupling of the theory. Here, we chose to use $\mu_u=750$ GeV like in \cite{Cacchio:2016qyh}.
We only take into account the quartic coupling terms from \cite{Cheng:2017tbn} and neglect the contributions of Yukawa and gauge couplings to the RGEs.
In Ref.~\cite{Cheng:2017tbn} it was shown how the parameter space is constrained in three cases where $\log_{10}(\Lambda/1\ {\rm GeV})=10,13,19$ if one uses LO unitarity and 2HDM stability.

\section{Results}
\label{sec:results}

While the boundedness-from-below constraints are trivial, we want to discuss the different unitarity constraints in the 2HDMW, before we consider higher scales and the effect of the theory constraints on the physical parameters for both, the 2HDMW and the MW model. Due to the large number of degrees of freedom, we present the direct comparison of \textit{model} parameters in the most cases. The results are also translated into \textit{physical} parameters, such as the scalar masses.

The contours in the figures presented below are the 100\% posterior probability regions. If we change the prior distribution of $\tan\beta$, for instance, replacing a flat $\log_{10}(\tan\beta)$ by a flat $\tan\beta$ prior, this will modify the shape of the posterior distributions (probably only slightly), but not the 100\% limits.

\subsection{Different unitarity constraints}

\begin{figure}
\begin{picture}(500,280)(0,0)
\centering
\put(-10,0){\includegraphics[width=\textwidth]{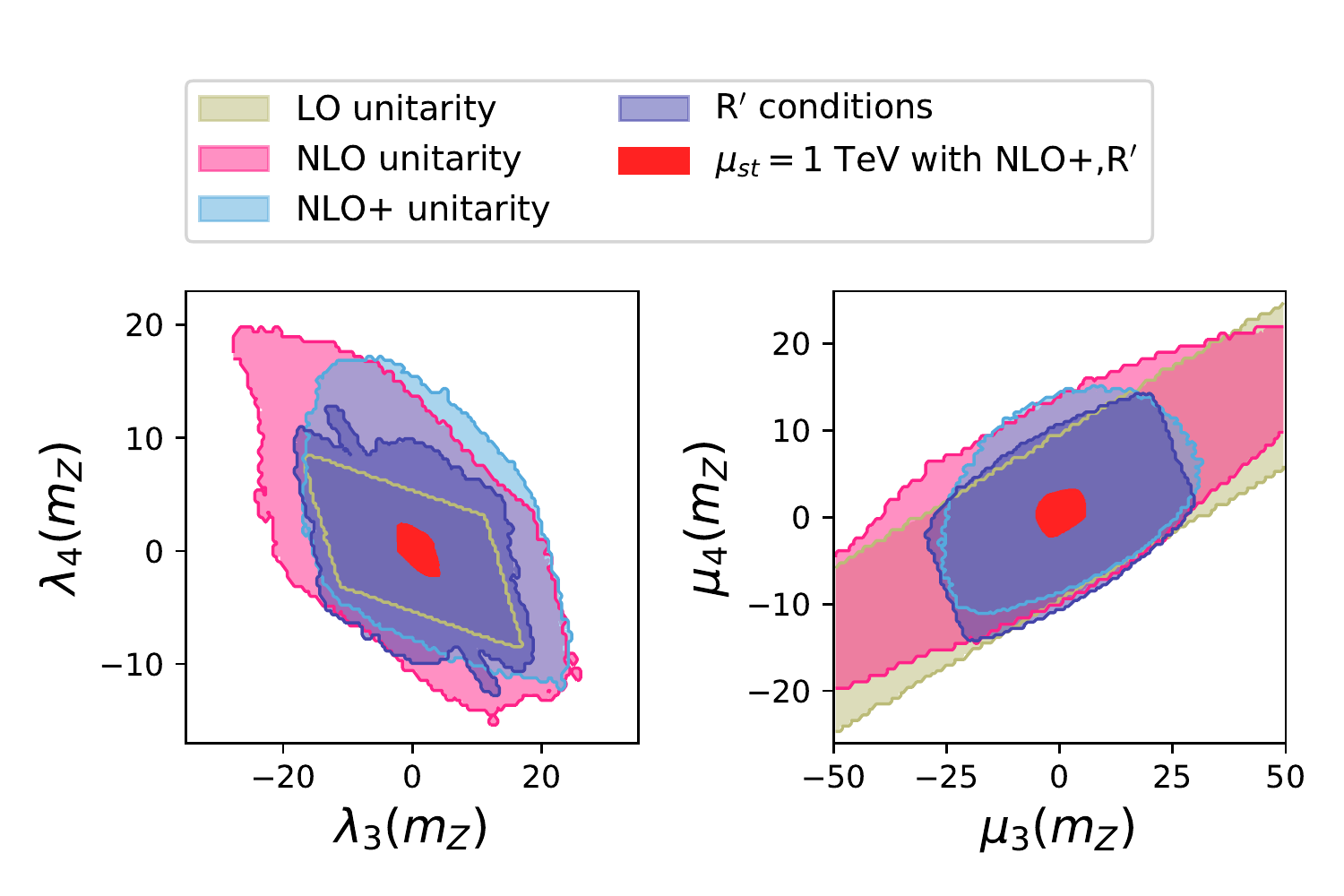}}
\put(400,230){\includegraphics[width=70pt]{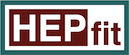}}
\end{picture}
\caption{Comparison of the different unitarity bounds in the $\lambda_4$ vs.~$\lambda_3$ and $\mu_4$ vs.~$\mu_3$ planes at the electroweak scale. The bounds shaded by colors other than red are obtained without renormalization-group running. Tree-level unitarity constrains the quartic couplings to the beige areas; the two sets of one-loop conditions NLO and NLO+ force the couplings to stay within the pink and light blue regions, respectively. The purple contour delimits the area compatible with the $R'$ conditions. The different unitarity bounds at the electroweak scale need to be compared to the regions with stable running including unitarity up to a scale of 1 TeV (red).}
  \label{fig:unitarity}
\end{figure}

In Figure \ref{fig:unitarity} we show the effects of LO, NLO and NLO+ criteria on the $\lambda_4$ vs.~$\lambda_3$ and $\mu_4$ vs.~$\mu_3$ planes as well as the impact of the $R'$ conditions explained in Section \ref{sec:unitarity} at the electroweak scale. Note that these bounds are calculated without running the renormalization-group equations, except for the red region. We observe that -- contrary to the 2HDM case, \textit{c.f.} Figure 2 of Ref.~\cite{Cacchio:2016qyh} -- the quartic couplings enjoy more freedom if we apply NLO(+) or the $R'$ criteria instead of the LO unitarity.
The reason for this is that the LO unitarity conditions only depend on few quartic couplings and disallow extreme values for them, while in the NLO(+) case, large quartic couplings can be compensated by tuning some of the other quartic couplings. Along the diagonal of the left hand panel of Figure \ref{fig:unitarity} we can observe the consequence of not applying the $R'$ criteria if the LO unitarity condition is accidentally small: In the small strip with $|\lambda_4+\lambda_3|\leq 0.01$ the quartic coupling $\lambda_4$ can be larger than $11$ in magnitude.
If we compare all sets of unitarity constraints with the region that is stable at least up to 1 TeV and compatible with NLO+ unitarity and the $R'$ conditions, we observe that the latter is a very strong bound. We would like to stress that we recommend to use the NLO(+) unitarity conditions only at scales significantly larger than the electroweak vev because beyond LO the quartic couplings are running couplings evaluated at an energy much larger than $v$.

\subsection{Combination of all theoretical constraints}

\begin{figure}
  \centering
\begin{picture}(500,290)(0,0)
\put(0,0){\includegraphics[width=\textwidth]{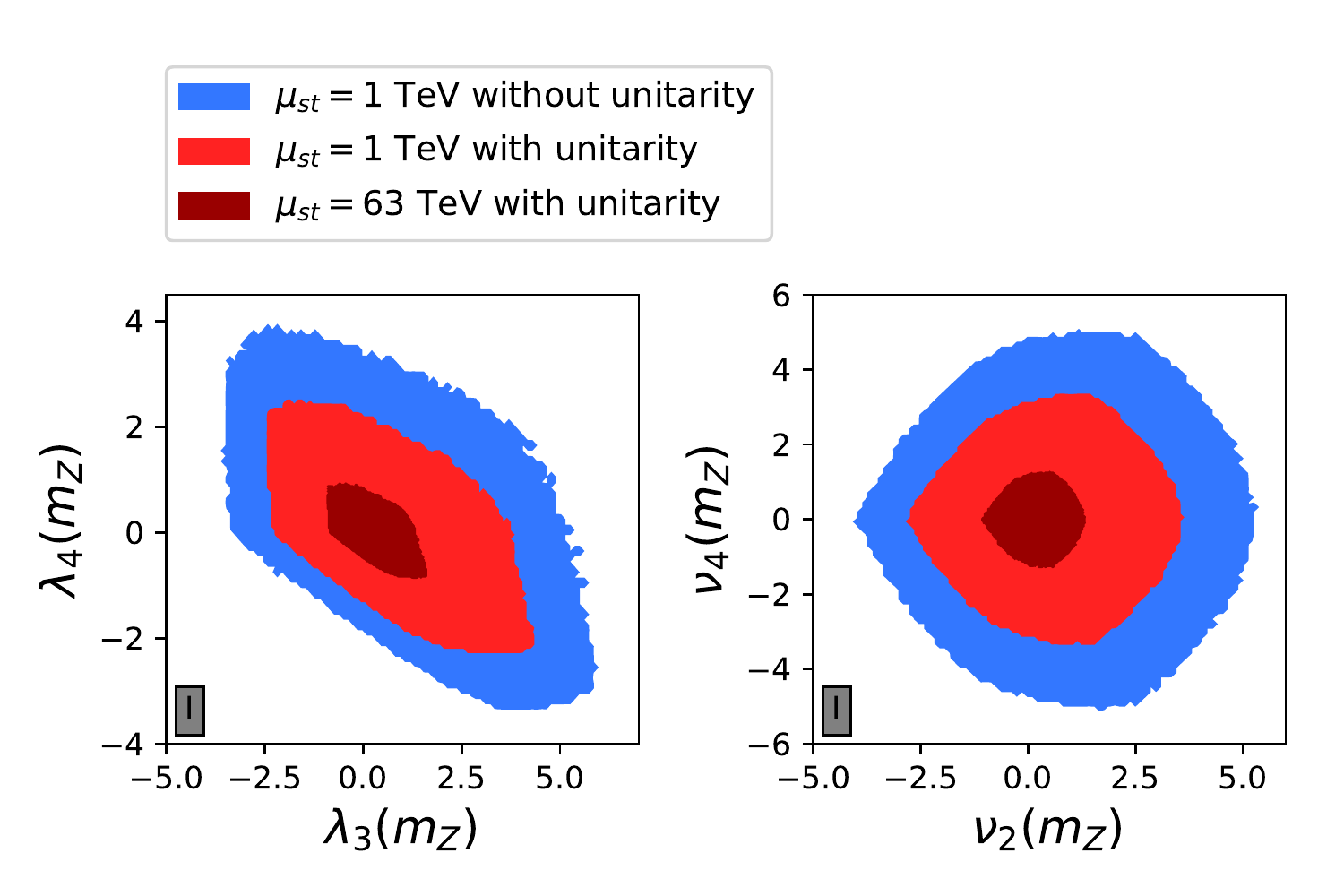}}
\put(300,240){\includegraphics[width=70pt]{HEPfitLogo.png}}
\end{picture}
 \caption{RG stability in the $\lambda_4$ vs.~$\lambda_3$ and $\nu_4$ vs.~$\nu_2$ planes of the 2HDMW of type I at the electroweak scale. The blue contours represent all scenarios that lead to a stable potential up to $1$ TeV without imposing any unitarity constraint, whereas NLO+ unitarity and $R'$ are added to the set of constraints for the red regions. The dark red region is compatible with all theory bounds and with a stable potential up to $63$ TeV.}
  \label{fig:running}
\end{figure}

In Figure \ref{fig:running} we illustrate the combination of the theory constraints with stability up to a certain scale in the $\lambda_4$ vs.~$\lambda_3$ and $\nu_4$ vs.~$\nu_2$ planes as representative examples of the 2HDMW. The limits obtained from the global fit to all quartic couplings of the 2HDMW and the MW limiting case can be found in Table \ref{tab:quarticlimits}. In this section, we analyze three different scenarios: In the first case, we run all quartic couplings to the stability scale of $\mu_{st}=1$ TeV, controlling at each iteration of the RG evolution if the potential is bounded from below and if all quartic couplings are in the perturbative regime, that is smaller than $4\pi$ in magnitude. We find that with these constraints, the absolute value of the quartic couplings at the electroweak scale cannot exceed limits between $3.3$ and $8.5$ ($1.7$ and $7.5$) without applying any unitarity bound to the 2HDMW (MW).
The has to be confronted with the second scenario, for which we add the NLO+ unitarity constraints as well as the $R'$ criteria at scales above $750$ GeV to the previous fit. The impact on the parameters is quite sizable: In Figure \ref{fig:running} we see that the allowed regions shrink by a factor of $1.5$ to $2$. The maximally allowed values for the quartic couplings range from $2.2$ to $5.7$ in the 2HDMW and from $1.3$ to $5.6$ in the MW, see Table \ref{tab:quarticlimits}.
For comparison, the theory-only upper limit on the quartic couplings of the 2HDM is $5.75$~\cite{Cacchio:2016qyh}.
Finally, we impose that the scalar potential with all discussed theory bounds is stable up to even higher scales $\Lambda$. Originally, we wanted to test high scales of $10^{4.8}$, $10^{7.6}$, $10^{12}$ and $10^{19}$ GeV, going in evenly spaced steps in the logarithm of $\log_{10} \Lambda$ towards the Planck scale, but our fitting set-up turned out to become unstable beyond $10^{4.8}$ GeV. If we choose $10^{4.8}$ GeV$\approx 63$ TeV as our high scale example, all parameters have to be between $-1.8$ and $2.2$ in the 2HDMW and between $-1.6$ and $2.2$ in the MW. Hence, the limits at $63$ TeV are stronger by a factor of about $2/5$ with respect to the ones obtained with stability at $1$ TeV. As a complete illustration, we arrange pairwise correlations of the bounds between all the couplings in Figs.~\ref{fig:triangle-plots1} and~\ref{fig:triangle-plots2} in the appendix.

In the MW limiting case the role of $\lambda_1$ is different, as it is the only parameter on which the mass of the SM-like Higgs depends; it is thus basically fixed by the Higgs mass measurements. Also, we do not impose any $\mathbb{Z}_2$ symmetry on the MW model, that is we treat $\nu_4$ as a free parameter.
The main difference between the limits on the quartic couplings in the 2HDMW and MW models is in how negative $\nu_1$ can be. Since $\lambda_1$ is fixed in the MW model, the positivity of the model limits the size of $\nu_1$. The Higgs trilinear coupling is sensitive to $\nu_1$ at the one-loop level. Thus larger values of $\nu_1$ are advantageous in trying to observe double-Higgs boson production at the LHC.

\begin{table}
{\footnotesize
\begin{tabular}{c|cccc|cccc|}
 & \multicolumn{4}{c|}{2HDMW limits} & \multicolumn{4}{c|}{MW limits}\\
Unitarity & -- & LO & NLO+,R' & NLO+,R' & -- & LO & NLO+,R' & NLO+,R' \\
$\mu_{st}$ & 1 TeV & 1 TeV & 1 TeV & 63 TeV & 1 TeV & 1 TeV & 1 TeV & 63 TeV \\
&\cellcolor[HTML]{3377ff} & &\cellcolor[HTML]{ff2222}  &\cellcolor[HTML]{990000} & & & &\\
\hline
$\lambda_1$ & [0, 3.9] & [0, 3.9] & [0, 2.7] & [0, 1.0] & \multicolumn{4}{c|}{$0.2585\pm 0.0007$} \\
$\lambda_2$ & [0, 3.9] & [0, 3.9] & [0, 2.7] & [0, 1.0] & \multicolumn{4}{c|}{--} \\
$\lambda_3$ & [-3.4, 5.8] & [-3.2, 5.5] & [-2.4, 4.2] & [-0.9, 1.6] & \multicolumn{4}{c|}{--} \\
$\lambda_4$ & [-3.3, 3.8] & [-3.2, 3.5] & [-2.2, 2.5] & [-0.9, 0.9] & \multicolumn{4}{c|}{--} \\
$\mu_1$ & [-5.5, 6.0] & [-5.3, 5.8] & [-3.8, 4.1] & [-1.5, 1.4] & [-5.3, 5.8] & [-5.3, 2.0] & [-3.6, 4.0] & [-1.4, 1.2] \\
$\mu_3$ & [-8.5, 7.8] & [-8.1, 7.7] & [-5.2, 5.7] & [-1.8, 2.2] & [-8.5, 7.5] & [0.0, 4.4] & [-5.1, 5.6] & [-1.6, 2.2] \\
$\mu_4$ & [-3.7, 4.9] & [-3.3, 4.8] & [-2.3, 3.2] & [-0.9, 1.2] & [-3.6, 4.8] & [-4.0, 2.3] & [-2.1, 3.1] & [-0.7, 1.2] \\
$\nu_1$ & [-4.7, 6.3] & [-4.5, 5.6] & [-3.1, 4.6] & [-1.2, 1.7] & [-1.7, 6.3] & [-1.2, 6.4] & [-1.3, 4.3] & [-0.8, 1.6] \\
$\nu_2$ & [-4.0, 5.2] & [-3.6, 5.0] & [-2.7, 3.5] & [-1.1, 1.3] & [-3.3, 5.1] & [-6.2, 6.4] & [-2.3, 3.4] & [-1.0, 1.3] \\
$\nu_4$ & [-5.0, 5.0] & [-4.8, 4.7] & [-3.3, 3.3] & [-1.3, 1.3] & [-4.6, 4.5] & [-7.6, 7.7] & [-2.9, 2.9] & [-1.1, 1.1] \\
$\omega_1$ & [-4.7, 6.3] & [-4.5, 6.0] & [-3.1, 4.5] & [-1.2, 1.7] & \multicolumn{4}{c|}{--} \\
$\omega_2$ & [-4.0, 5.2] & [-3.9, 5.1] & [-2.8, 3.5] & [-1.1, 1.3] & \multicolumn{4}{c|}{--} \\
$\omega_4$ & [-4.9, 4.9] & [-4.8, 4.7] & [-3.2, 3.3] & [-1.3, 1.3] & \multicolumn{4}{c|}{--} \\
\hline
$m_A-m_H$ [GeV] & [-390, 440] & [-340,400] & [-340, 360] & [-210, 230] & \multicolumn{4}{c|}{--} \\
$m_R-m_I$ [GeV] & [-320, 370] & [-280,330] & [-260, 310] & [-170, 190] & [-250, 300] & [-100, 230] & [-180, 250] & [-150, 180] \\
\end{tabular}
}
\caption{Limits on the quartic couplings and two mass differences with different assumptions. The second to fourth columns contain the 2HDMW results. Note that $\nu_4$ ($\omega_4$) is only non-zero in the case(s) of the type I, IIu (IId) 2HDMW. Columns five to seven contain the results of the MW limiting case. In this case, $\lambda_1=m_h^2/v^2$.}
\label{tab:quarticlimits}
\end{table}

Comparing our results with those of Ref.~\cite{Cheng:2017tbn}, we find that our allowed ranges for the quartic couplings assuming stability and NLO unitarity up to 63 TeV are more or less of the same size as previous limits using LO unitarity and no MW positivity up to $2\cdot 10^4$ TeV.

\begin{figure}
  \centering
\begin{picture}(500,290)(0,0)
\put(0,0){\includegraphics[width=\textwidth]{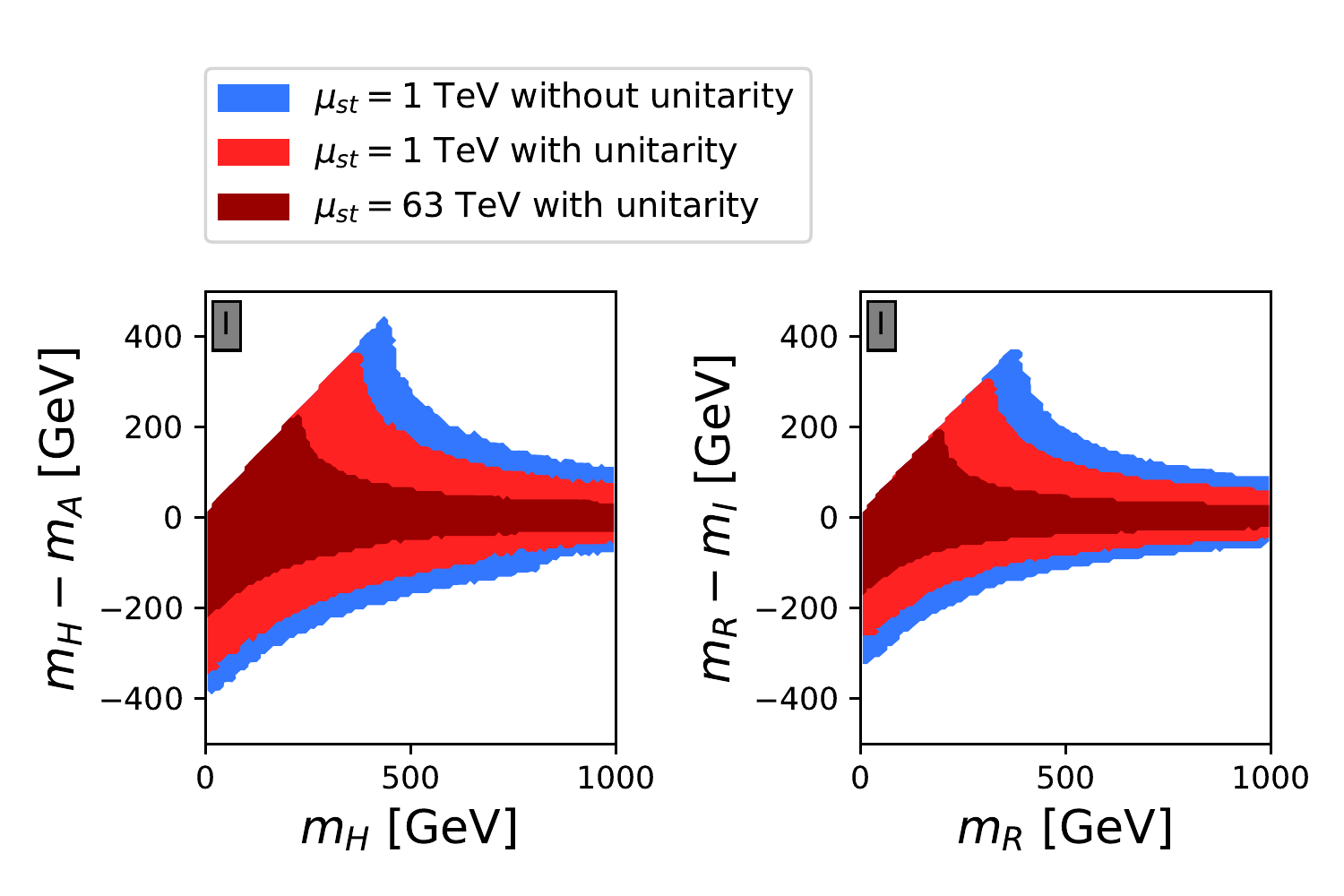}}
\put(300,240){\includegraphics[width=70pt]{HEPfitLogo.png}}
\end{picture}
 \caption{Comparison of different stability scales in the $m_H-m_A$ vs.~$m_H$ and $m_R-m_I$ vs.~$m_R$ planes of the Type I 2HDMW. For the color code we refer to Fig.~\ref{fig:running}.}
  \label{fig:masses}
\end{figure}

The limits on the quartic couplings can be translated into bounds on the physical model parameters. Like in the 2HDM we observe strong restrictions of the differences between $m_H^2$ and $m_A^2$ \cite{Cacchio:2016qyh, Chowdhury:2017aav}, but also $m_I^2$ cannot deviate very much from $m_R^2$, see Figure \ref{fig:masses}. The former mass square difference depends on the values of the $\lambda_i$, while $m_R^2-m_I^2$ is proportional to $\nu_2 c^2_\beta+\omega_2 s^2_\beta$, see eq.~\eqref{eq:mRsqminusmIsq}.
The linear dependence of the upper limit of the mass splittings for light $m_H$ and $m_R$ comes from requiring both masses to be positive. This feature does not appear in the analogous limit in Figure 6 of~\cite{Cacchio:2016qyh} because the mass splittings in that figure are plotted against the mass of a third Higgs boson.
Fits to the mass differences $m_A-m_H$ and $m_R-m_I$ in the three mentioned scenarios yield upper bounds between $440$ and $170$ GeV in the 2HDMW and between $300$ and $150$ GeV in the MW model, see the last two rows of Table \ref{tab:quarticlimits}. 
Just like for the 2HDMW, the theory-only limit on the mass splitting in the 2HDM is 360~GeV~\cite{Cacchio:2016qyh}.
Even if Fig.~\ref{fig:running} and \ref{fig:masses} were obtained for the Type I 2HDMW, the 2HDMW limits in Table \ref{tab:quarticlimits} hold for all three types, only that either $\nu_4$ or $\omega_4$ have to be set to zero, depending on the type.

\section{Conclusions}
\label{sec:conclusions}

We have studied the NLO unitarity bounds on the 2HDMW, which extends the scalar sector of 2HDM with an additional color octet scalar. Although less constraining than the LO unitarity bounds at the electro-weak scale, the NLO unitarity constraints become stronger when running up to higher scales greater than $1$~TeV. However, compared with the MW model which is the limiting case of 2HDMW, the common quartic couplings, i.e. $\mu$'s and $\nu$'s, are allowed for larger ranges under these constraints.

In addition, we have derived a set of necessary conditions to bound the 2HDMW potential from below for the first time. These conditions constrain most of the quartic couplings except a few. They are also applicable to the limiting case of the MW model.

Finally, we have combined all theoretical constraints and found limits of the couplings assuming stability at different scales. Requiring a stable potential at a higher scale favors smaller mass differences between pairs of neutral scalars, such as $m_A-m_H$ and $m_R-m_I$.

The next obvious step would be a combination with experimental constraints, for which our publicly available \HEPfit implementation could be used.

\section*{Acknowledgments}

We thank C.~Murgui, A.~Pich and G.~Valencia for fruitful discussions. We thank the INFN Roma Tre Cluster, where most of the fits were performed.
The work of OE was supported by the Agencia Estatal de Investigaci\'{o}n (AEI, ES) and the European Regional Development Fund (ERDF, EU) [Grants No.~FPA2014-53631-C2-1-P, FPA2017-84445-P and SEV-2014-0398].
The work of CM was supported by the United States Department of Energy under Grant Contract {DE-SC0012704}.

\appendix
\section*{Appendix}
\label{sec:appendix}

\begin{figure}
\centering
\begin{picture}(500,580)(0,0)
\put(50,0){\includegraphics[width=350pt,trim=0 0 395 15,clip=true]{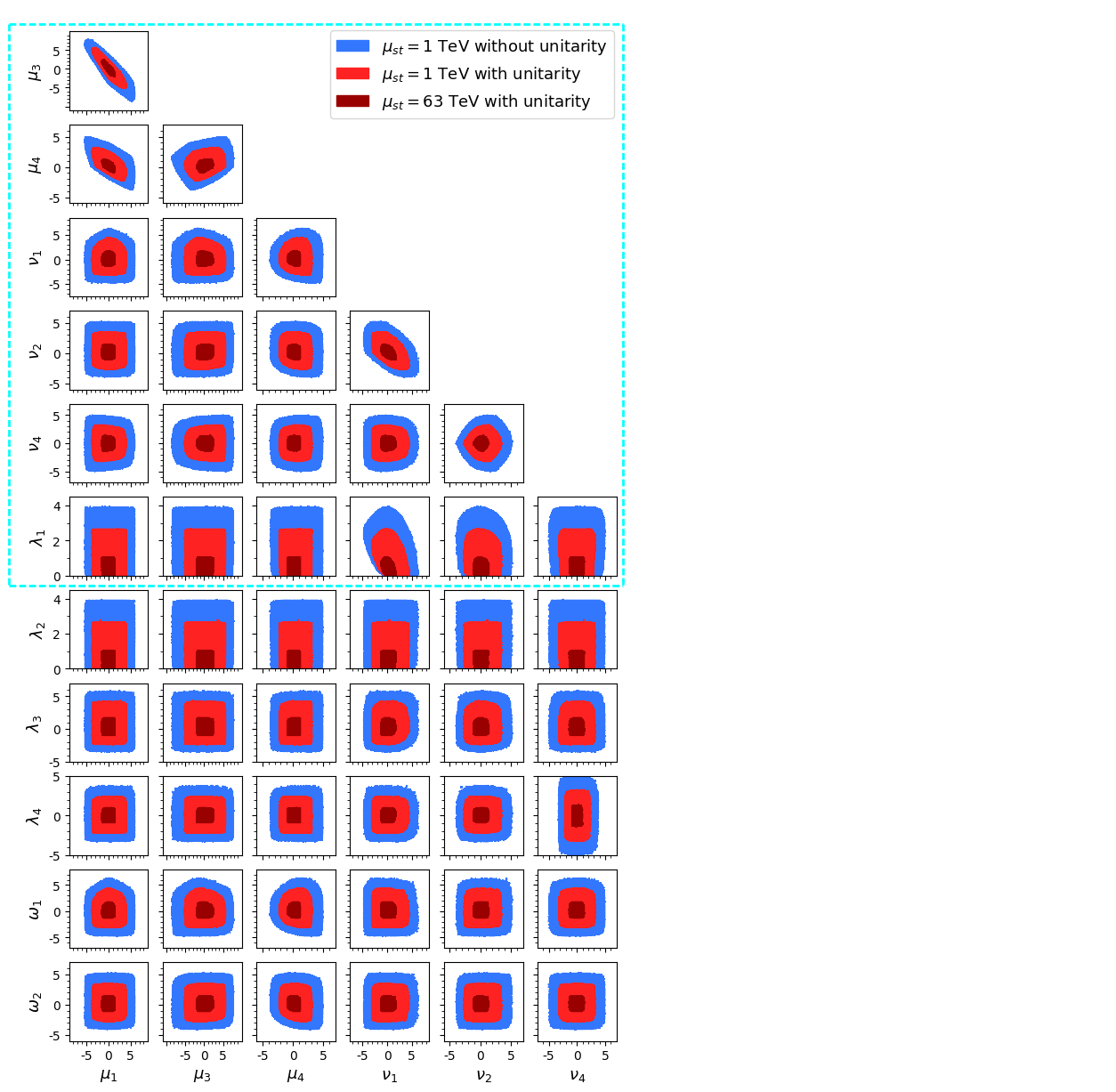}}
\put(300,440){\includegraphics[width=70pt]{HEPfitLogo.png}}
\end{picture}
\caption{Pairwise correlations of bounds between all the couplings at different scales. The colored scheme of representing different constraints is the same as that in Fig.~\ref{fig:running}. The cyan dashed box contains the MW parameters.}
\label{fig:triangle-plots1}
\end{figure}

\begin{figure}
\centering
\begin{picture}(500,350)(0,0)
\put(70,0){\includegraphics[width=380pt,trim=350 0 0 320,clip=true]{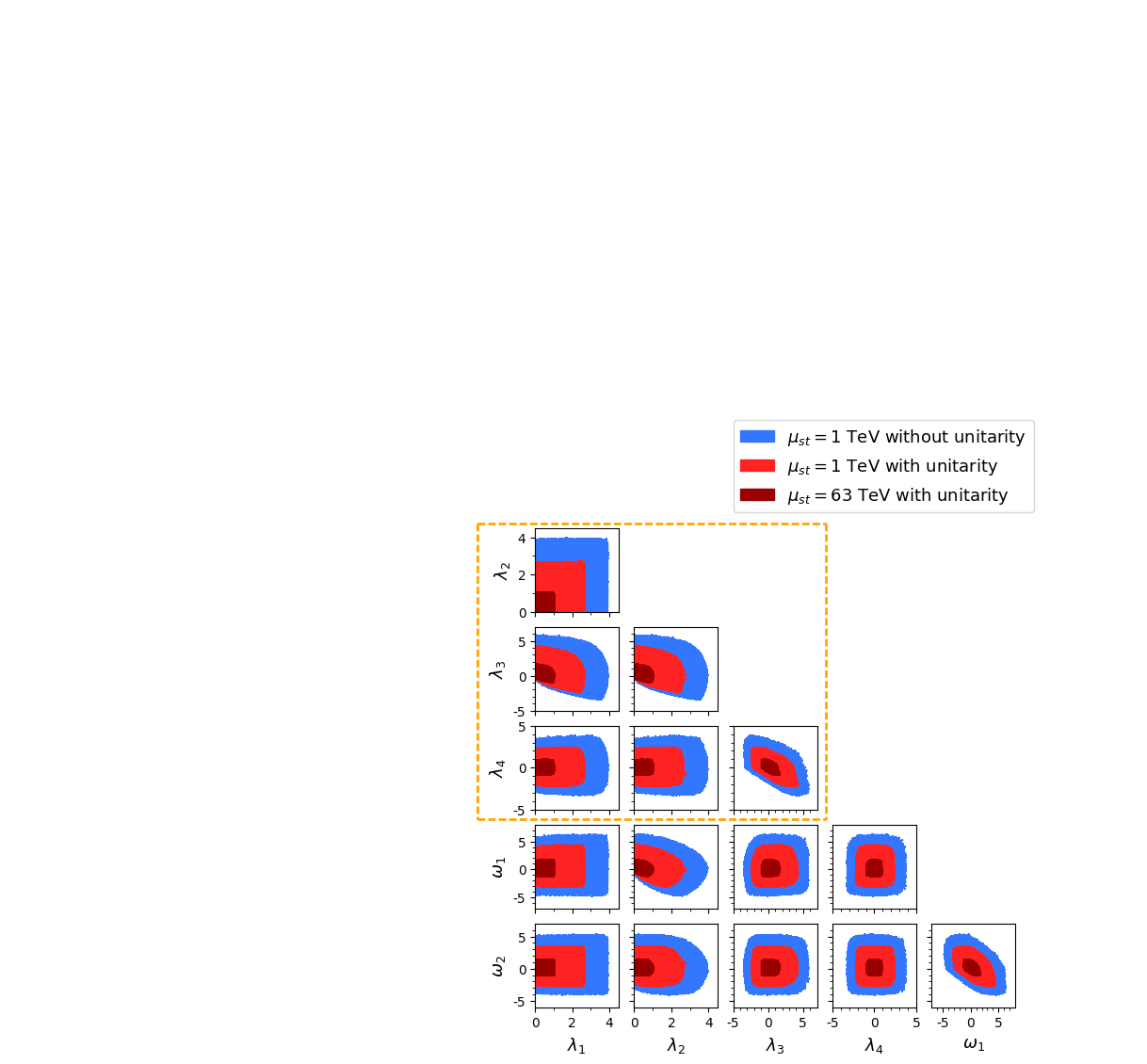}}
\put(300,240){\includegraphics[width=70pt]{HEPfitLogo.png}}
\end{picture}
\caption{Pairwise correlation of bounds between all the couplings at different scales. The colored scheme of representing different constraints is the same as that in Fig.~\ref{fig:running}. The orange dashed box contains the 2HDM parameters.}
\label{fig:triangle-plots2}
\end{figure}

For the sake of completeness we give the two-dimensional correlations of the 2HDMW in Fig.~\ref{fig:triangle-plots1} and Fig.~\ref{fig:triangle-plots2}. The former also contains the MW subset, while the latter includes the 2HDM correlations.

\bibliographystyle{apsrev4-1}
\bibliography{references}

\end{document}